\documentclass[journal=jpclcd,manuscript=article]{achemso}

\usepackage{chemformula,amssymb} 
\usepackage[T1]{fontenc} 
\usepackage{aas_macros}

\author{Alexandre Faure}
\affiliation{Universit\'e Grenoble Alpes, CNRS, IPAG, 38000 Grenoble, France}
\email{alexandre.faure@univ-grenoble-alpes.fr}
\author{Fran\c{c}ois Lique}
\affiliation{LOMC - UMR 6294, Normandie Universit\'e, Universit\'e du Havre and CNRS, 25 rue Philippe Lebon, BP 1123 - 76 063 Le Havre cedex, France}
\email{francois.lique@univ-lehavre.fr}
\author{Anthony J. Remijan}
\affiliation{National Radio Astronomy Observatory, 520
  Edgemont Rd., Charlottesville, VA 22903, USA}
\email{aremijan@nrao.edu}

\title[Weak Maser Action of Interstellar Methanimine] {Collisional
  Excitation and Weak Maser Action of Interstellar Methanimine}

\abbreviations{IR,NMR,UV}

\begin{document}

\begin{tocentry}

\includegraphics[width=5.0cm, angle = 0 ]{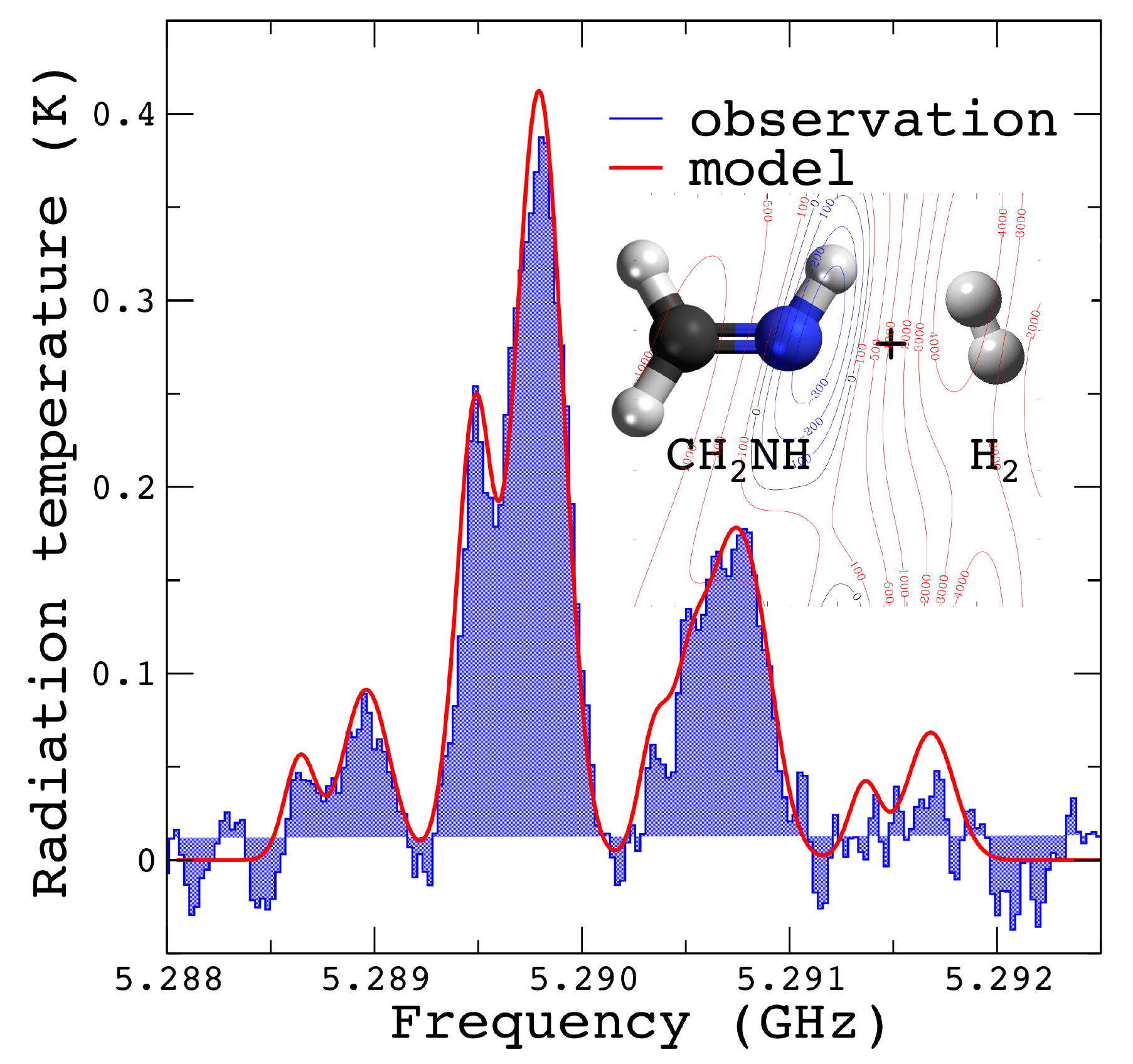}

\end{tocentry}

\begin{abstract}

The inelastic cross sections and rate coefficients for the rotational
excitation of methanimine (CH$_2$NH) by cold H$_2$ have been
determined quantum mechanically based on a new highly correlated
five-dimensional potential energy surface. This surface was fitted to
more than 60 000 {\it ab initio} points with a root mean square error
of $\sim$ 1-2~cm$^{-1}$ in the region of the potential well whose
depth is -374.0~cm$^{-1}$. The rotationally inelastic rate
coefficients have been combined with spectroscopic data to generate
the non-equilibrium spectrum of CH$_2$NH towards the interstellar
molecular cloud Sgr~B2(N). The transition $1_{10}\to 1_{11}$ at
5.29~GHz is found to be inverted and the predicted weak maser emission
spectrum is in excellent agreement with new observations performed
with the 100-meter Green Bank Telescope.


\end{abstract}


Complex organics are observed throughout the Universe, from Solar
System objects to stars and galaxies. In the interstellar medium
(ISM), more than 200 different (mostly organic) molecules have been
detected, including biomolecule precursors such as formamide
(NH$_2$CHO) and glycolaldehyde (CH$_2$OHCHO) (see e.g. Belloche {\em
  et al.}  \cite{belloche13}). Imines are of special interest because
they are possible precursors of amino acids \cite{danger11}, which
have so far eluded detection in the ISM \cite{snyder05}. Amino acids
are the building blocks of proteins and the study of their precursors
may shed light about the origin of life in the Universe.

Methanimine (CH$_2$NH), the simplest imine, was discovered in the ISM
in 1973 towards the molecular cloud Sagittarius~B2
(Sgr~B2)\cite{godfrey73}.  It was firmly identified thanks to the
(hyperfine) multiplet structure of the $1_{10}\to 1_{11}$ rotational
line at 5.29~GHz. Since then, CH$_2$NH has been detected in many
galactic and extragalactic sources (see Ref.~\citenum{dore12} for a
review). The formation of methanimine in the ISM is, however, matter
of debate. Ion-molecule reactions are the dominant gas-phase processes
at low temperatures and the formation of CH$_2$NH might occur through
the reaction CH$_3^+$ + NH$_2 \to $CH$_2$NH$_2^+$ + H followed by the
dissociative recombination of CH$_2$NH$_2^+$ with an electron to
produce CH$_2$NH \cite{turner99}. Neutral-neutral reactions involving
radicals such as NH + CH$_3$ are other possible gas-phase formation
pathways \cite{suzuki16}. Finally, hydrogenation of solid HCN is a
possible grain-surface route to CH$_3$NH$_2$ where CH$_2$NH is a
stable intermediate, as demonstrated experimentally \cite{theule11}.

In order to elucidate the chemistry of CH$_2$NH in the ISM, it is
first crucial to observe multiple transitions in a large frequency
range. This has been made possible through many laboratory studies of
the rotational spectrum of methanimine (see Ref.~\citenum{dore12} and
references therein). As a result, CH$_2$NH has been detected towards
Sgr~B2 from centimeter \cite{godfrey73} to submillimeter wavelengths
\cite{persson14}. The identification of many transitions is, however,
not sufficient to derive accurate abundances because interstellar
spectra are generally not at local thermodynamic equilibrium
(LTE). Indeed, densities (i.e. pressures) in the ISM are such that the
frequency of collisions is neither negligible nor large enough to
maintain thermodynamic equilibrium.  In such conditions, deriving
column densities (i.e. abundances) requires to solve simultaneously
the radiative transfer equation and a set of statistical equilibrium
equations for the molecular levels. This in turn requires the
availability of the state-to-state rate coefficients for collisional
(de-)excitation. These collisional energy transfer rate coefficients
are extremely difficult to measure in the laboratory and radiative
transfer models rely almost exclusively on theoretical estimates
\cite{roueff13}. In the case of CH$_2$NH, the only theoretical study
was done for electron-impact excitation \cite{wang15}. In the
``dense'' molecular ISM, however, the dominant colliders are hydrogen
molecules (H$_2$). In this Letter, we report the first potentiel
energy surface (PES) for CH$_2$NH--H$_2$ based on $\sim$ 66,000
high-level {\it ab initio} points. This PES was employed in quantum
close-coupling calculations to determine rotationally inelastic cross
sections in the energy range 2$-$200~cm$^{-1}$, from which rate
coefficients were deduced in the temperature range 5$-$30~K. In
addition, the spectrum of CH$_2$NH towards Sgr~B2(N) at 5.29~GHz has
been generated using radiative transfer calculations and it has been
compared to new observations performed with the 100-m Robert C. Byrd
Green Bank Telescope (GBT).

\begin{figure}
  \centering{\includegraphics[width=7.0cm, angle = 0 ]{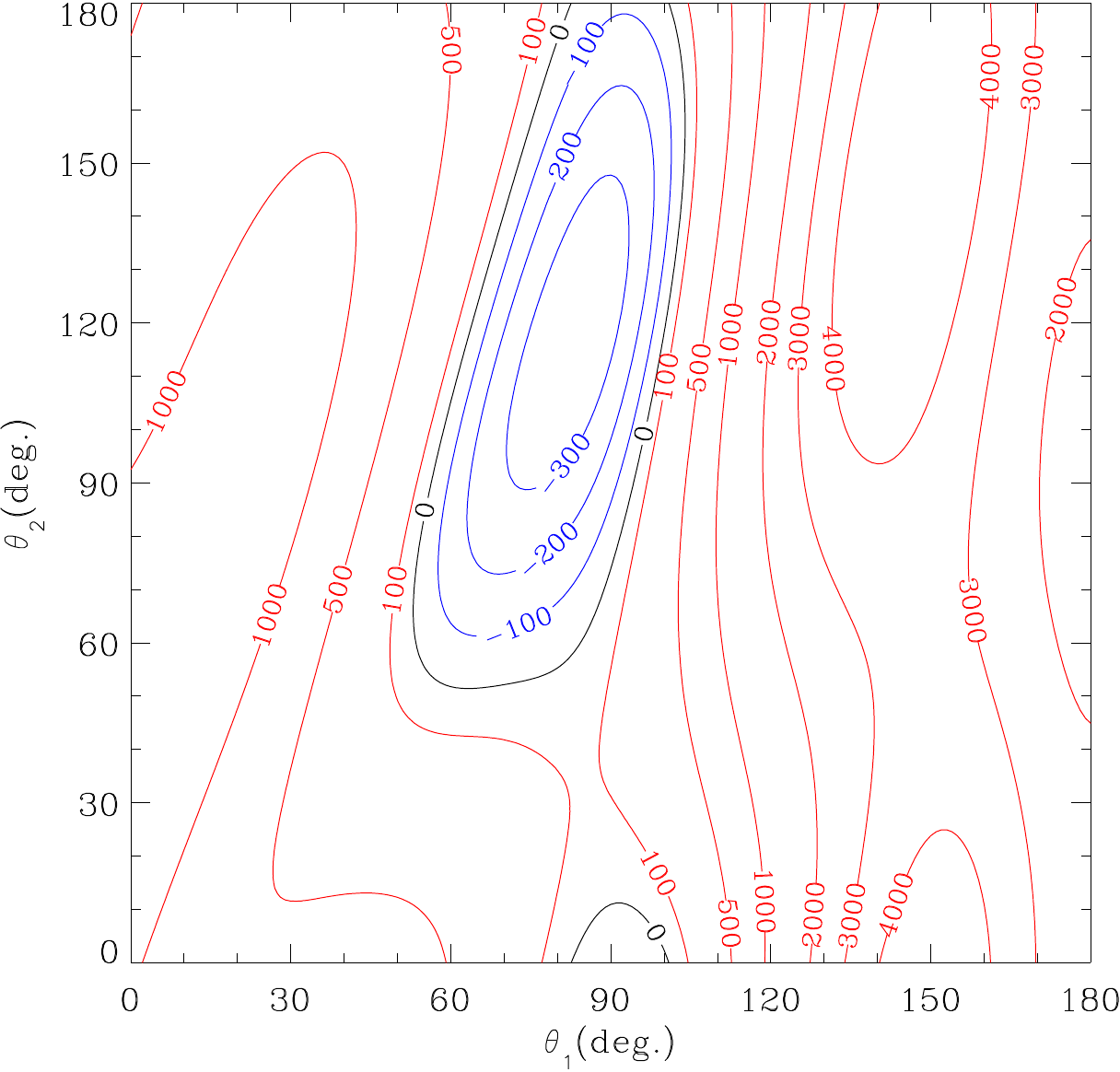}}
  \centering{\includegraphics[width=7.0cm, angle = 0 ]{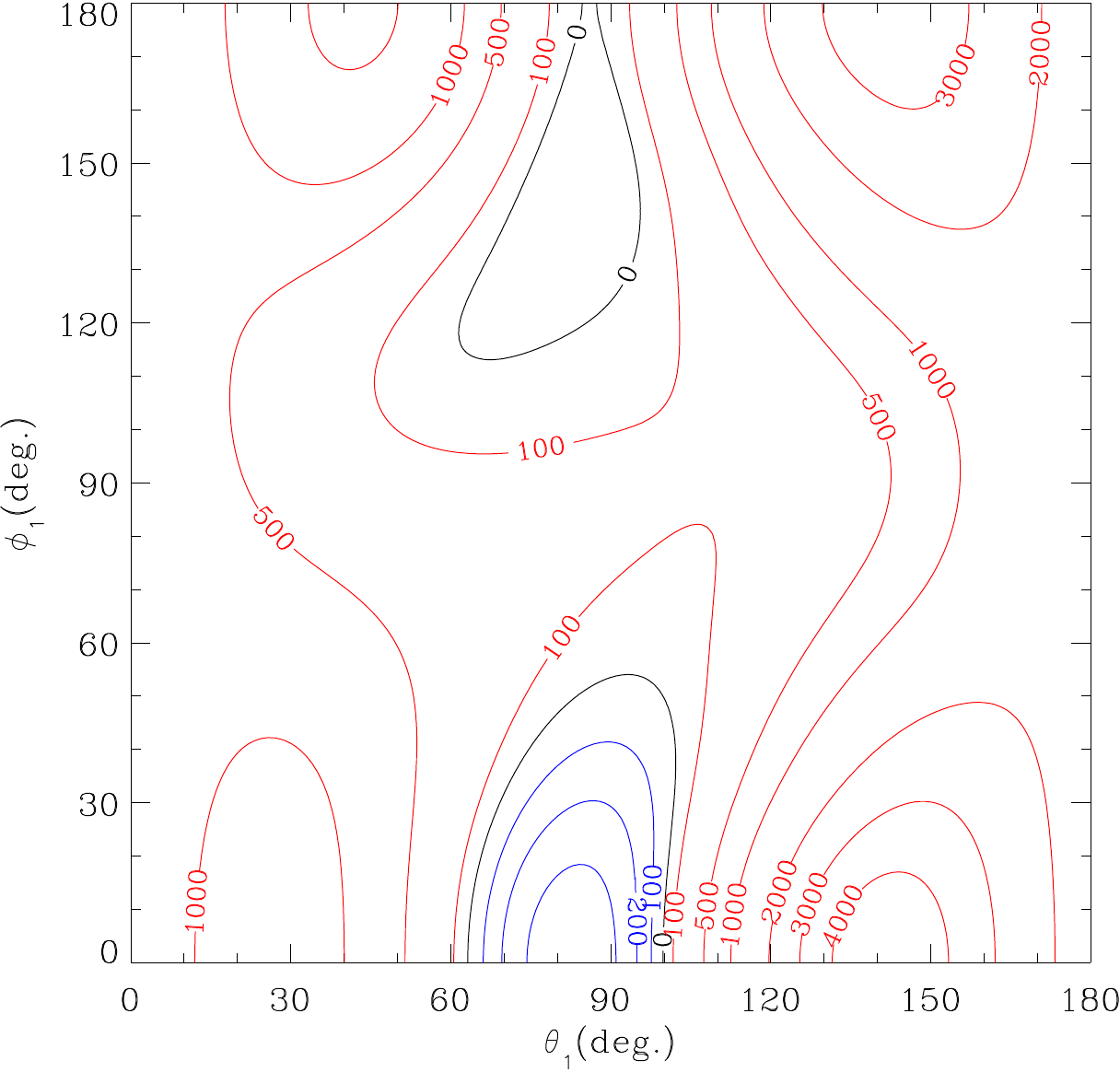}}
\caption{Potential energy cuts of the 5D CH$_2$NH$-$H$_2$ PES. The
  intermolecular distance is fixed at $R=5.77$~$a_0$ and
  $\phi_2=0^{\circ}$, corresponding to the equilibrium values of the
  complex. Left panel: $\phi_1=0^{\circ}$. Right panel:
  $\theta_2=117^{\circ}$. The minimum in the two plots is
  -374.0~cm$^{-1}$ at $\theta_1=83^{\circ}$. Energy contours in
  cm$^{-1}$.}
\label{fig:pes}
\end{figure}

The PES of the CH$_2$NH$-$H$_2$ complex was computed in a body-fixed
frame centered at the center of mass of CH$_2$NH, as defined in
Ref.~\citenum{phillips94}. Here the $z$- and $x$-axes correspond to
the $a$- and $b$- principal inertia axes of the molecule,
respectively, so that CH$_2$NH lies in the $(x, z)$ plane. The
intermolecular vector ${\bf R}$ connects the centres of mass of the
CH$_2$NH and H$_2$ molecules. The $\theta_1$ and $\phi_1$ angles
define the rotation of vector ${\bf R}$ relative to the CH$_2$NH
body-fixed axis system. The rotation of the H$_2$ molecule is defined
by angles $\theta_2$ and $\phi_2$. The monomers were assumed to be
rigid\footnote{It was shown recently for the CO$-$H$_2$ system that
  the use of ground vibrational state geometries in rigid-rotor PES is
  an excellent approximation to a full-dimensional approach
  \cite{faure16}, which here would involve 15-dimensional
  calculations.} with geometries corresponding to the ground
vibrational state: $r({\rm CN})$=2.406~a$_0$, $r({\rm
  NH})$=1.929~a$_0$, $r({\rm CH})$=2.06~a$_0$, $\widehat{\rm
  HNC}$=110.4$^{\circ}$,$\widehat{\rm
  HCH}$=117.0$^{\circ}$,$\widehat{\rm NCH_{\rm cis}}$=125.1$^{\circ}$
from the experimental results of Ref.~\citenum{pearson77} and $r({\rm
  HH})=1.44874$~a$_0$ from the calculations of Ref.~\citenum{bubin03}.
{\it Ab initio} calculations for the CH$_2$NH$-$H$_2$ complex in its
ground electronic state were carried out at the explicitly correlated
coupled cluster with single, double and perturbative triple
excitations [CCSD(T)-F12a]\cite{knizia09} level of theory with
augmented correlation-consistent triple zeta (aVTZ) \cite{dunning89}
basis set using the \textsf{MOLPRO} 2012 package \cite{MOLPRO}. Basis
set superposition error (BSSE) was corrected using the Boys and
Bernardi counterpoise scheme\cite{boys70}. The size inconsistency
caused by the F12 triple energy correction was corrected by
subtracting from the {\it ab initio} interaction energies the
asymptotic interaction energy at $R=1000$~$a_0$ which is 7.4796
cm$^{-1}$ for all relative orientations. The performance of the
CCSD(T)-F12a method was assessed by comparison with standard CCSD(T)
results using double-, triple- and quadruple-zeta basis sets
extrapolated to the complete basis set (CBS) limit. We have found that
CCSD(T)-F12 calculations are within 1$-$2~cm$^{-1}$ of
CCSD(T)/CBS(D,T,Q) results in the minimum region of the PES.
The calculations were carried out for a large random grid of angular
orientations: for each value of $R$, varied from 4.0 to 15.0~$a_0$,
the energies of 3000 mutual orientations were calculated. The energies
were least-squares fitted to an angular expansion of 251 terms (see
computational details). The global minimum of the (fitted) PES occurs
at $R=5.77$~$a_0$, $\theta_1=83^{\circ}, \phi_1=0^{\circ},
\theta_2=117^{\circ}, \phi_2=0^{\circ}$ with a dissociation energy of
$D_e=374.0$~cm$^{-1}$. In this planar geometry, the complex has an
approximate T-shape with H$_2$ almost perpendicular to the C=N
bond. In Fig.~\ref{fig:pes}, we present two contour plots of the
CH$_2$NH$-$H$_2$ (fitted) PES in this region. We can observe that the
PES is highly anisotropic at this distance, with only a small
attractive region where $\theta_1\sim 80^{\circ}$.

\begin{figure}
  \centering{\includegraphics[width=7.0cm, angle = -90 ]{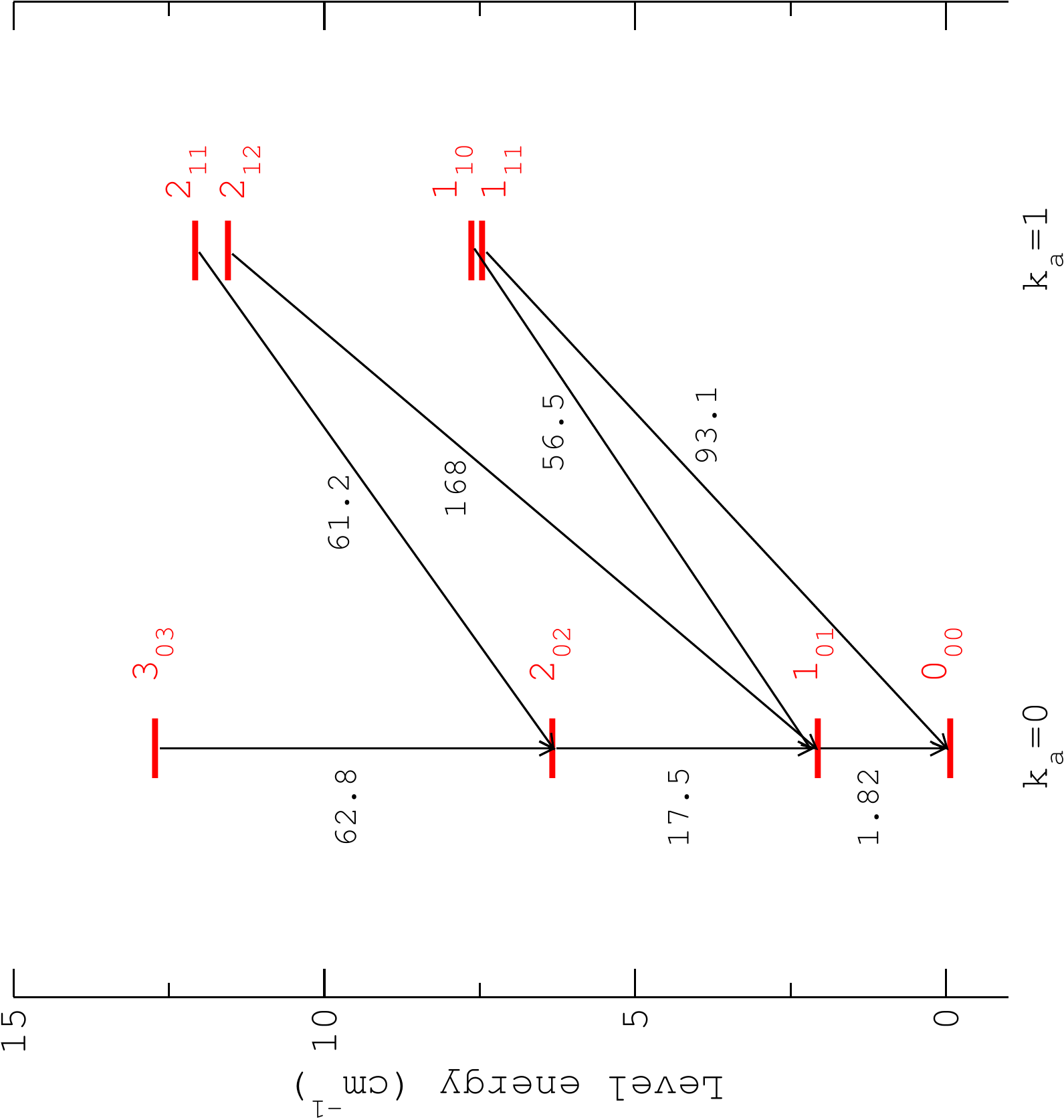}}
  \caption{Lowest 8 levels of CH$_2$NH, with energies taken from the
    CDMS catalog \cite{muller05}. For each level, the strongest
    radiative transitions are indicated by arrows and the numbers give
    the corresponding Einstein coefficient (in units of
    10$^{-6}$~s$^{-1}$) taken from Ref.~\citenum{muller05}. The
    spectrum consists of $a$-type ($\Delta k_a=0$) and $b$-type
    ($\Delta k_a=\pm 1$) transitions with corresponding dipole moments
    $\mu_a=1.340$~D and $\mu_b=1.446$~D.}
\label{fig:lev}
\end{figure}

\begin{figure}
  \centering{\includegraphics[width=7.0cm, angle = -90 ]{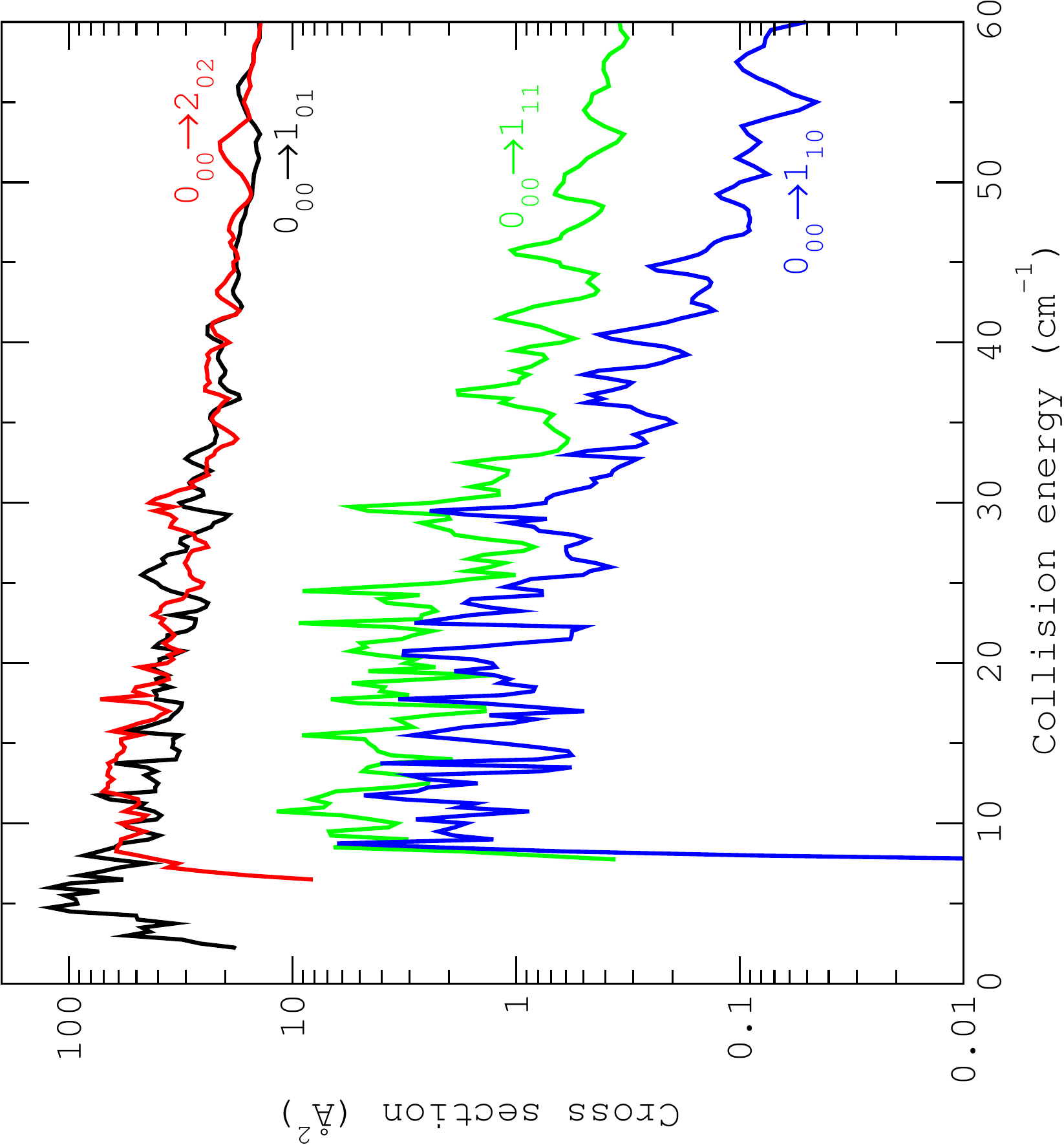}}
  \centering{\includegraphics[width=7.0cm, angle = -90 ]{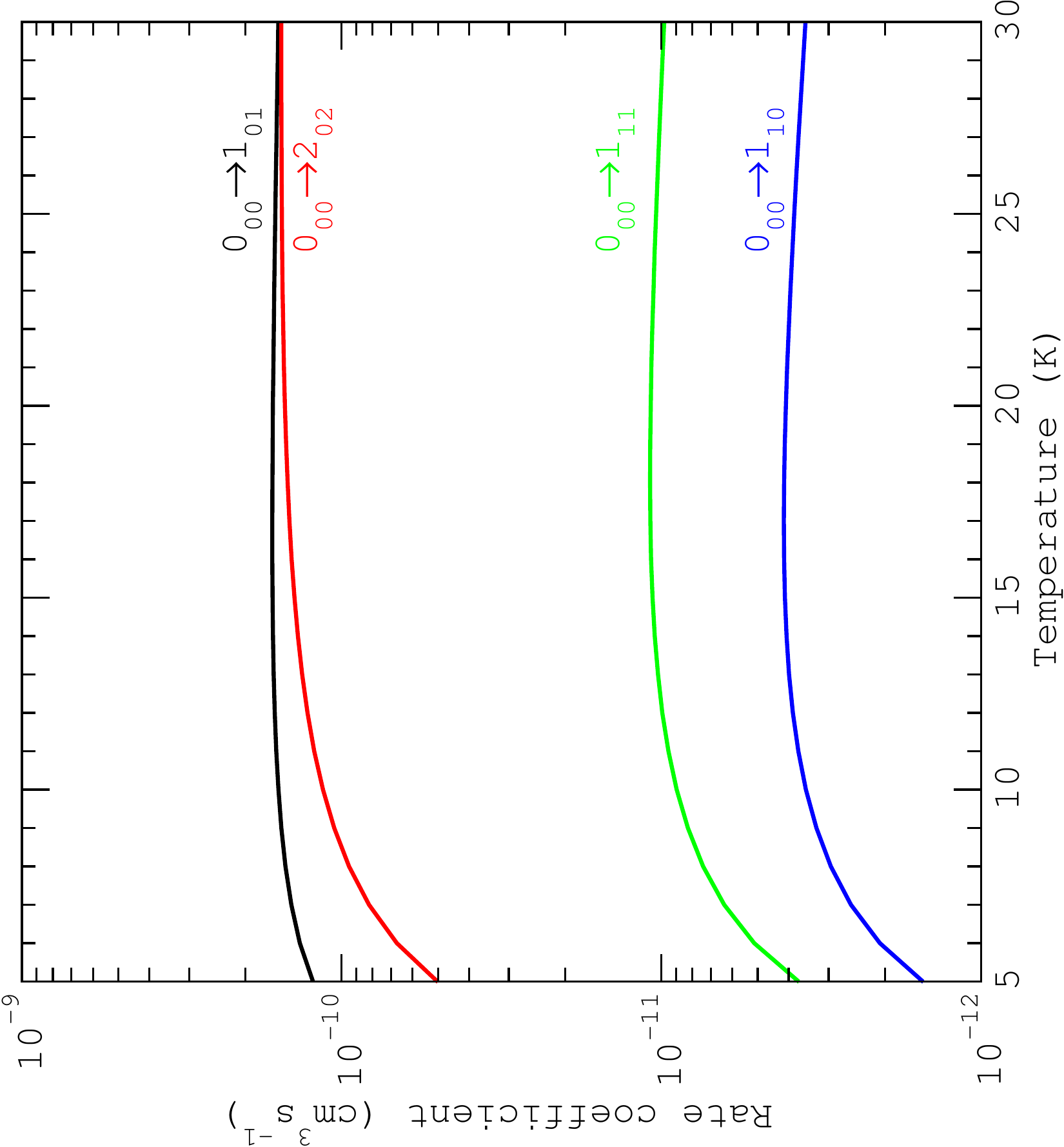}}
\caption{Cross sections (left panel) and rate coefficients (right
  panel) for excitation out of the $0_{00}$ level to levels $1_{01}$,
  $2_{02}$, $1_{11}$ and $1_{10}$ due to para-H$_2(j_2=0)$
  collisions.}
\label{fig:xs}
\end{figure}

The CH$_2$NH$-$H$_2$ PES was employed in quantum-mechanical scattering
calculations to compute rotationally inelastic cross sections. We used
the (time independent) close-coupling formalism for collisions of an
asymmetric top rigid rotor and a linear rigid rotor, as described in
Ref.~\citenum{phillips95}. All calculations were performed with the
OpenMP extension\footnote{See
  http://ipag.osug.fr/$\sim$faurea/molscat/index.html} of the version
14 of the \textsf{MOLSCAT} code \cite{molscat}. The coupled
differential equations were solved using the hybrid modified
log-derivative Airy propagator from 3 to a maximum of 200~$a_0$ (at
low energy) with a step size lower than 0.25~$a_0$. The rotational
constants of CH$_2$NH were fixed at their experimental
values\cite{muller05} $A_0=6.54490$~cm$^{-1}$, $B_0=1.15552$~cm$^{-1}$
and $C_0=0.979062$~cm$^{-1}$. The rotational constant of H$_2$ was
taken as $B_0=59.322$~cm$^{-1}$. The rotational energy levels of
CH$_2$NH are labeled by three quantum numbers: the angular momentum
$j_1$ and the pseudo-quantum numbers $k_a$ and $k_c$ which correspond
to the projection of $j_1$ along the axis of the smallest and largest
moments of inertia, respectively. The lowest 8 levels of CH$_2$NH are
displayed in Fig.~\ref{fig:lev}.
The rotational energy levels of H$_2$ are designated by $j_2$. Due to
the large well depth in the CH$_2$NH$-$H$_2$ PES (374~cm$^{-1}$) and
the relatively dense rotational spectrum of CH$_2$NH, close-coupling
calculations were highly demanding both in terms of memory and CPU
time. In practice, calculations were performed for total energies
between 2 and 200~cm$^{-1}$, with an energy grid adjusted to properly
sample the low-energy resonances. The projectile H$_2$ was restricted
to para-H$_2$($j_2=0$), which is the dominant form of H$_2$ in the
cold interstellar medium \cite{troscompt09,faure13}. The basis set
included levels $j_1=0-12$ and $j_2=0, 2$ with rotational energies
(CH$_2$NH plus H$_2$) less than a threshold
\texttt{EMAX}=600~cm$^{-1}$. This cutoff was necessary to limit the
number of coupled-channels below 6,000 while preserving convergence to
within $\sim$10\% for all transitions among the first 15 rotational
levels of CH$_2$NH, i.e. up to the level $2_{20}$ at
28.3~cm$^{-1}$. Rate coefficients were obtained in the temperature
range 5--30~K by integrating the cross sections over Maxwell-Boltzmann
distributions of relative velocities.
Cross sections and rate coefficients for the lowest 4 transitions out
of the ground state $0_{00}$ are presented in Fig.~\ref{fig:xs} as
function of collisional energy and temperature, respectively. Many
resonances are observed in the cross sections, as expected from the
deep potential well. We can also notice that the favored transitions
are $0_{00}\to 1_{01}$ and $0_{00}\to 2_{02}$, corresponding to the
usual propensity rule $\Delta j_1=1, 2$ and $\Delta k_a=0$. In
addition, for the $k_a=1$ doublet, the lower level is found to be
favored. This propensity was also observed for higher $k_a=1$ doublets
and this corresponds to $j_1$ being preferentially oriented along the
direction of the greatest moment of inertia (the $c-$axis). This
result is observed in other molecules and is responsible, in
particular, of the anti-inversion of the doublet $1_{10}-1_{11}$ in
interstellar H$_2$CO \cite{troscompt09}. Here, however, because both
collisional and radiative transitions between the $k_a=1$ and $k_a=0$
ladders are allowed (in contrast to H$_2$CO), the above propensity is
in competition with inter-ladder transitions. In particular the
spontaneous decay rate from the $1_{11}$ to $0_{00}$ level is about
1.6 times that from the $1_{10}$ to the $1_{01}$ level (see
Fig.~\ref{fig:lev}), which favors inversion of the doublet, as in the
case of methyl formate (HCOOCH$_3$) \cite{faure14}. The strong
emission of the intrinsically weak $1_{10}\to 1_{11}$ line, as first
detected by Godfrey {\em et al.}  \cite{godfrey73}, was thus
attributed to a population inversion rather than an anomalously high
abundance \cite{turner91}. The lack of collisional data, however, has
precluded so far any definite conclusion. Using our rate coefficients,
the critical density\footnote{The critical density defines the density
  at which the collisional rate out of the upper state of a transition
  equals the spontaneous radiative rate.} of the upper level $1_{10}$
can be estimated as $\sim 1.5\times 10^5$~cm$^{-3}$ which means that
non-LTE effects are expected at densities in the range
$10^3-10^7$~cm$^{-3}$. However, only non-LTE radiative transfer
calculations can predict the actual rotational populations of CH$_2$NH
for a given set of physical conditions, as illustrated below.

\begin{figure}
  \centering{\includegraphics[width=7.0cm, angle = -90 ]{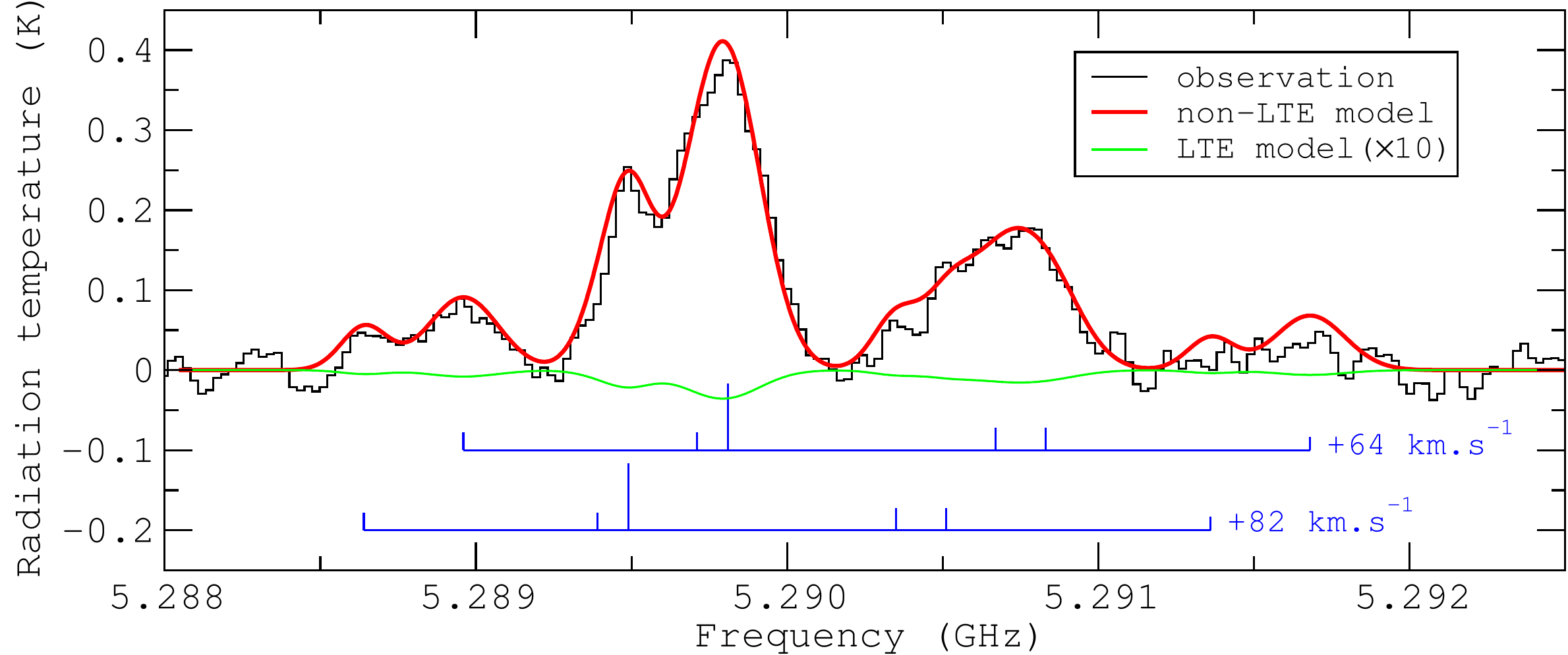}}
\caption{Observational and model spectra of methanimine $1_{10}\to
  1_{11}$ transition at 5.29~GHz towards Sgr~B2(N). Relative
  intensities of the (partially resolved) hyperfine structure in the
  optically thin limit are shown at the bottom. The nominal source
  velocity is +64~km.s$^{-1}$. A second velocity component is resolved
  at +82~km.s$^{-1}$. The non-LTE model predicts a population
  inversion with a negative excitation temperature
  $T_{ex}=-0.48$~K. The LTE model spectrum has been multiplied by a
  factor of 10 to show it more clearly.}
\label{spec}
\end{figure}

The star forming region Sgr~B2 is an exceptionally massive cloud
complex in the Galactic Center, where shocks and supersonic turbulence
are ubiquitous. Towards the north core of Sgr~B2 (Sgr~B2(N)), the
continuum emission is produced by a complex region of ionized gas
(so-called HII region). A detailed description of this source can be
found in Ref.~\citenum{corby15}. We focus below on the $1_{10}\to
1_{11}$ transition of CH$_2$NH at 5.29~GHz. This line was first
observed in 1973 by Godfrey {\em et al.} \cite{godfrey73} towards
Sgr~B2 with the 3.8$'$ beam (at 5.29~GHz) of the Parkes 64-meter
telescope. New observations towards Sgr~B2(N) were performed with the
2.3$'$ beam of the GBT. The multiplet was detected with a high
signal-to-noise ratio, as displayed in Fig.~\ref{spec}. The transition
frequencies have been shifted to the laboratory rest frame using the
nominal source velocity of +64~km.s$^{-1}$. Weaker transitions with a
velocity component of 82~km.s$^{-1}$ (-0.32~MHz) are also observed, as
found in several other molecules with the GBT \cite{remijan08}. These
two velocity components correspond to two molecular clouds lying
within the GBT field of view along the line of sight towards
Sgr~B2(N).

Radiative transfer calculations were performed with the non-LTE
\textsf{RADEX} program \cite{vandertak07} based on the escape
probability formalism for a uniform sphere. The code was employed to
compute the line excitation temperature ($T^i_{ex}$) and opacity
($\tau_i$) for each hyperfine transition $i=1, 6$ of the $1_{10}\to
1_{11}$ line. The two velocity components were computed
separately. The hyperfine energy levels and radiative rates were taken
from the CDMS catalog \cite{muller05}. The hyperfine-resolved rate
coefficients were obtained from the rotational rate coefficients using
the statistical approximation, which has been shown to be reliable at
low opacities \cite{faure12}. The kinetic temperature and density of
H$_2$ were fixed at $T$=30~K at $n_{\rm H_2}=10^4$~cm$^{-3}$,
respectively, for each velocity component. These values, derived by
Faure {\em et al.}  \cite{faure14} from their modeling of the weak
HCOOCH$_3$ masers, correspond to a relatively cold and dilute gas
surrounding the north core, in the foreground of the HII region. The
line widths were taken as 15 and 10~km.s$^{-1}$ for the low- and
high-velocity components, respectively \cite{corby15}. The column
density of CH$_2$NH was adjusted for each velocity component to best
reproduce the observational spectrum. The solution of the radiative
transfer equation was expressed as the radiation temperature, as in
Ref.~\citenum{faure14}:
\begin{equation}
  \Delta T^*_R(\nu)=[J_{\nu}(T_{ex})-J_{\nu}(T_{CMB})-T_c(\nu)](1-e^{-\tau(\nu)}),
    \label{eq:ta}
\end{equation}
where $J_{\nu}(T)=(h\nu/k_B)/(e^{h\nu/k_BT}-1)$, $\nu$ is the
frequency of the transition (5.29~GHz), $T_{cmb}=2.725$~K is the
temperature of the cosmic microwave background and $T_c(\nu)$ is the
main beam background continuum temperature of Sgr~B2(N), as measured
by the GBT, which is $\sim$36~K at 5.29~GHz. The total opacity
$\tau(\nu)$ in Eq.~\ref{eq:ta} is the sum over the two velocity
components and the 6 hyperfine transitions, assuming individual
opacities are gaussian. The resulting composite line (with 8 resolved
components) is compared to the observation in Fig.~\ref{spec} where a
remarkable agreement is obtained. The best fit was found for CH$_2$NH
column densities of $2.5 \pm 0.5\times 10^{14}$ and $1.0 \pm 0.2\times
10^{14}$~cm$^{-2}$ for the low- and high-velocity components,
respectively. As expected, the $1_{10}\to 1_{11}$ transition is
inverted with an excitation temperature of -0.48~K (equal for each
hyperfine transition within 0.01~K). We have checked that this
population inversion occurs for H$_2$ densities in the range $6\times
10^2-6\times 10^6$~cm$^{-3}$, as anticipated, meaning this is a very
robust phenomenon. We have also plotted the emergent spectrum in LTE
conditions with the same column densities. A very faint absorption
(the spectrum has been multiplied by a factor of 10) is predicted in
this case, demonstrating that the weak inversion, by amplifying the
strong background continuum radiation, is essential to produce the
emission of the otherwise undetectable 5.29~GHz line.

The derived CH$_2$NH column densities correspond to low fractional
abundances relative to H$_2$ ($f({\rm H_2}) \lesssim 10^{-10}$,
assuming a H$_2$ column density of $3\times 10^{24}$~cm$^{-2}$ as in
Ref.~\citenum{nummelin00}). They are in particular lower than the
value derived by Halfen {\em et al.}  \cite{halfen13} ($9.1\pm 4.4
\times 10^{14}$~cm$^{-2}$) from a spectral survey of emission lines at
higher frequencies ($>$70~GHz), which probe another extended, warmer
and denser, component.
Methanimine is also detected in a compact component, the so-called
``hot core'' of Sgr~B2(N) where the column density is $\sim
10^{18}$~cm$^{-2}$ \cite{belloche13}. A more complete treatment is now
required to model all detected lines, in absorption and emission, in
order to improve the determination of the abundances and physical
conditions in both the extended cool and warm components. This will be
presented in a future dedicated work.

\section{Computational details}

The analytical expansion for an asymmetric-top-linear molecular system
can be written as \cite{phillips94}:
\begin{equation}
  V(R,\theta_1,\phi_1,\theta_2,\phi_2)=\sum
  v_{l_1m_1l_2l}(R)t_{l_1m_1l_2l}(\theta_1,\phi_1,\theta_2,\phi_2),
\end{equation}
where the functions $t_{l_1m_1l_2l}(\theta_1,\phi_1,\theta_2,\phi_)$
are products of spherical harmonics, as given explicitly in Eqs.~(5-6)
of Ref.~\citenum{valiron08}. The indices $l_1$, $m_1$ and $l_2$ refer
to the $(\theta_1, \phi_1)$ and $(\theta_2, \phi_2)$ dependence of the
PES, whereas $l$ runs from 0 to the sum of $l_1$ and $l_2$. The
expansion coefficients $v_{l_1m_1l_2l}(R)$ were obtained through a
least-squares fit on the random grid of 3000 angular geometries at
each intermolecular separation (22 radial grid points in the range
$R=4-15~a_0$). All anisotropies up to $l_1=14$, $l_2=6$ and $l=20$
were included, resulting in 3054 expansion functions. We then selected
only significant terms using a Monte Carlo error estimator (defined in
Ref.~\citenum{rist12}), resulting in a final set of 251 expansion
functions with anisotropies up to $l_1=13$, $l_2=6$ and $l=16$. The
root mean square was found to be lower than 2~cm$^{-1}$ in the
long-range and minimum regions of the PES ($R>5.75~a_0$). In these
regions, the mean error on the expansion coefficients
$v_{l_1m_1l_2l}(R)$ is smaller than 1.25~cm$^{-1}$. A cubic spline
radial interpolation of these coefficients was finally employed over
whole grid of intermolecular distances and it was smoothly
extrapolated (using exponential and power laws at short- and
long-range, respectively) in order to provide continuous radial
coefficients adapted to quantum close-coupling calculations.

\begin{acknowledgement}

This work was supported by the French program Physique et Chimie du
Milieu Interstellaire (PCMI) funded by the Conseil National de la
Recherche Scientifique (CNRS) and Centre National d'Etudes Spatiales
(CNES). F.L. acknowledges the Institut Universitaire de France for
financial support. We thank Joanna Corby for stimulating discussions
initiating this work. The National Radio Astronomy Observatory is a
facility of the National Science Foundation operated under cooperative
agreement by Associated Universities, Inc. The Green Bank Observatory
is a facility of the National Science Foundation operated under
cooperative agreement by Associated Universities, Inc.

\end{acknowledgement}




  

\bibliography{faure2705}

\end{document}